# Evaluating the exfoliation of two-dimensional materials with a Green's function surface model

Yipeng An,[1,*] Yusheng Hou,[2] Shijing Gong,[3,4] Ruqian Wu,[2,†] Chuanxi Zhao,[5] Tianxing Wang,[1] Zhaoyong Jiao,[1] Heyan Wang,[1] and Wuming Liu[6,‡]

[1] *School of Physics, International United Henan Key Laboratory of Boron Chemistry and Advanced Energy Materials, Henan Normal University, Xinxiang 453007, China*
[2] *Department of Physics and Astronomy, University of California, Irvine, California 92697, USA*
[3] *Key Laboratory of Polar Materials and Devices (MOE), School of Physics and Electronic Science, East China Normal University, Shanghai 200062, China*
[4] *Collaborative Innovation Center of Extreme Optics, Shanxi University, Taiyuan 030006, China*
[5] *Siyuan Laboratory, Guangdong Provincial Engineering Technology Research Center of Vacuum Coating Technologies and New Energy Materials, Department of Physics, Jinan University, Guangzhou 510632, China*
[6] *Beijing National Laboratory for Condensed Matter Physics, Institute of Physics, Chinese Academy of Sciences, Beijing 100190, China*

Previous methods for the evaluation of the exfoliation of two-dimensional (2D) layered materials have drawbacks in computational efficiency and are unable to describe cases with semi-infinite substrates. Based on a Green's function surface (GFS) model, here we develop an approach to efficiently determine the tendency of exfoliation of 2D materials from their bulk crystals or semi-infinite substrates. By constructing appropriate surface configurations, we may calculate the exfoliation energy more precisely and quickly than the traditional way with the slab model. Furthermore, the GFS approach can provide angle-resolved photoemission spectroscopy (ARPES) of surface systems for direct comparison with experimental data. Our findings indicate that the GFS approach is powerful for studies of 2D materials and various surface problems.

## I. INTRODUCTION

Since the successful preparation of graphene [1], lots of 2D monolayers (MLs) have been predicted, and many of them have been synthesized. Typical 2D MLs such as h-BN [2], silicene [3, 4], phosphorene [5–8], borophene [9–12], transition metal sulfides [13–16], MXene [17, 18], stanine [19], antimonene [20], g-$C_3N_4$[21, 22], and ferromagnets (e.g., $CrX_3$, $Fe_3GeTe_2$, and $GdAg_2$) [23–28], have been extensively explored. The versatile properties of these 2D materials are promising for diverse applications such as superconductors, spintronic or topotronic films, and electrodes in lithium-sulfur batteries. At the present stage, the ways to prepare 2D materials can be generally classified into two categories: top-down approaches and bottom-up approaches. The first way is to cleave sheets of layered materials, using various exfoliation techniques [1, 29–31]. The other way is to grow them on appropriate substrates through chemical reactions with specific precursors, including the chemical vapor deposition (CVD) [32] and wet-chemical syntheses [33].

The most critical step to get high quality free-standing 2D materials is to exfoliate them from their van der Waals(vdW) layered crystals or from substrates. A crucial criterion for the feasibility of exfoliating 2D materials is the exfoliation energy (EE), which is the key to determine if the material can be separated easily in experiments. EEs are typically calculated through density functional theory (DFT) approaches, mostly using the slab model with a finite number of atomic layers (see Supplemental Material [34]). Approximations such as neglecting possible lattice relaxation of exfoliated layers and the remaining interaction through thin slabs, make results less reliable. Expansion of slab thickness or supercell size makes calculations expensive. Recently, it was shown that the EE of a given material is equal to the energy difference between a single isolated layer and its bulk (per layer) [35], which can somewhat reduce the computational cost. It is desired to find a general and efficient approach to more reliably determine EEs of 2D materials, especially from semi-infinite substrates.

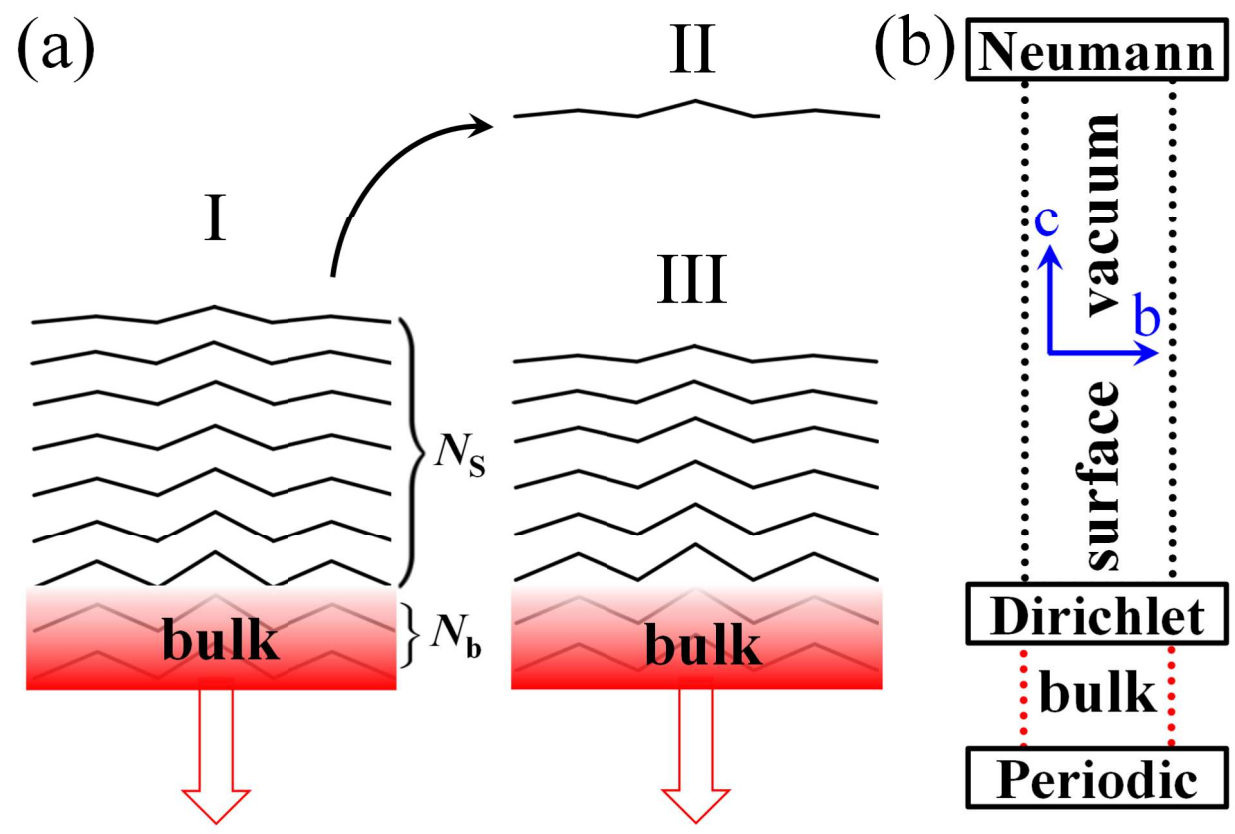

FIG. 1. (a) Schematic of the Green's function surface model for calculating the exfoliation energies. I (III) is the original (remaining) structure, and II is the exfoliated layer. $N_b$ and $N_S$ refer to the numbers of atomic layers in the bulk and surface regions, respectively. (b) Boundary conditions of the Green's function surface model.

* ypan@htu.edu.cn
† wur@uci.edu
‡ wmliu@iphy.ac.cn

In this paper, we propose that the exfoliation energies of 2D monolayers or few-layers, either from their bulk crystals or from semi-infinite substrates, can be evaluated by a Green's function approach [36, 37]. As exfoliating a certain 2D material from its bulk or a thick sample, one may mimic the process using a periodic unit-cell (see Fig. 1(a)). Compared to the slab method, this greatly reduces the number of atoms in DFT calculations and can be employed to evaluate the exfoliation of more complex cases and thick substrates.

## II. METHOD

All calculations of this work are performed by using the density functional theory and the Green's function methods, as implemented in the Atomistix Toolkit code [38–41]. The exchange and correlation effects of electrons are described with the spin-polarized GGA-PBE functional [42, 43]. The core electrons of all atoms are represented by the optimized Norm-Conserving Vanderbilt pseudo-potentials [44]. Linear combinations of atomic orbitals (LCAO) basis sets are employed to expand the wave-functions of valence electrons. A real-space grid density that is equivalent to a plane-wave kinetic energy cutoff of 100 Ha is adopted. The 21×21×1 Monkhorst-Pack $k$-point grids are used for the Brillouin zone sampling for the slab structures and the Green's function surfaces (GFSs). For the bulk region of the GFS structures, the $k$-point grids are 21×21×100, to make sure that electronic structures of the bulk and surface regions match well. The total energy tolerance and residual force on each atom are less than $10^{-6}$ eV and 0.01 eV/Å in the geometry relaxation, respectively. A vdW correction (DFT-D2) is also employed in the calculations [45].

## III. RESULTS AND DISCUSSION

For the traditional slab model, the EE is defined as the difference between the ground-state energy of a $N$-layer thick slab, the exfoliated layer and the remaining ($N$-1)-layer slab [46–48]. The central bulk region should be thick enough to obtain a convergent result. For the GFS model (see Fig. 1(a)), the surface region contains a few layers of a 2D material (or plus several substrate layers), and the bulk region described by a periodic unit-cell is semi-infinite along the negative $c$ axis. The electronic structures across the bulk-surface boundary are matched using the Green's functions. The Hamiltonian matrix of this semi-infinite system can be described as

$$\mathbf{H}^{\mathrm{KS}} = \begin{pmatrix} \ddots & \vdots & \vdots & \vdots & \vdots \\ \cdots & \mathbf{V}_{\mathrm{BB}}^{\dagger} & \mathbf{H}_{\mathrm{B}} & \mathbf{V}_{\mathrm{BB}} & 0 \\ \cdots & 0 & \mathbf{V}_{\mathrm{BB}}^{\dagger} & \mathbf{H}_{\mathrm{B}} & \mathbf{V}_{\mathrm{BS}} \\ \cdots & 0 & 0 & \mathbf{V}_{\mathrm{BS}}^{\dagger} & \mathbf{H}_{\mathrm{S}} \end{pmatrix}, \tag{1}$$

where $\mathbf{H}_{\mathrm{S}}$ and $\mathbf{H}_{\mathrm{B}}$ denote the Hamiltonians of the surface and semi-infinite bulk regions, respectively. $\mathbf{V}_{\mathrm{BB}}$ and $\mathbf{V}_{\mathrm{BS}}$ are the coupling matrices which describe the interaction between the principal layers of the semi-infinite bulk part and between the bulk-surface interface, respectively. The electronic structures of these surface systems can be determined by solving the Kohn-Sham and Poisson equations within the three different boundary conditions (BCs) along the $c$ axis (see Fig. 1(b)) [37, 49]. The bulk region is periodic towards the negative $c$ axis, and the bulk-surface interface adopts a Dirichlet BC. The Neumann BC is used at the top vacuum boundary, but we found that the Dirichlet BC gives almost the same results. These boundary conditions are often employed in studies of semi-infinite surface configurations or two-probe systems [37, 49]. As shown in Fig. 1(a), the EE is described as the energy difference between the original structure (I), the exfoliated layer (II), and the remaining structure (III). Namely, the exfoliation energy (per unit area) $E_{\mathrm{exf}}$ is defined as

$$E_{\mathrm{exf}} = (E_{\mathrm{II}} + E_{\mathrm{III}} - E_{\mathrm{I}})/A, \tag{2}$$

where $A$ refers to the in-plane area of the surface unit-cell. More details are given in the Supplemental Material [34].

To examine the reliability of this GFS model for evaluating the exfoliation energies, we first apply it to some well-known 2D materials: graphene, h-BN, transition metal sulfides, and chromium trihalides. We systematically study two cases of exfoliations from their bulk crystals and from a substrate. For the first case, we calculate the EEs of monolayer ($E_{\mathrm{exf}}(1)$) and few-layer ($E_{\mathrm{exf}}(\mathrm{n})$), respectively. For the second case, we take the exfoliation of graphene from a Ni(111) surface as an example, which is often adopted to prepare graphene.

Fig. 2(a) shows the monolayer EEs of graphene, h-BN, and $MoS_2$, calculated by the slab and GFS methods. To overcome the size effect, we examine the EEs with a $N_{\mathrm{S}}$ from 5 to 9 layers. These two methods give almost the same results, indicating the reliability of the GFS model. For instance, the calculated EE of graphene is 35 meV/Å$^2$, which is close to the experiment-based value 29 meV/Å$^2$ [50] and better than the plane-wave result (21 meV/Å$^2$) [35]. The required computation time for the exfoliation energy of the GFS method is a few times shorter than that of the slab method (see Fig. 2(a)).

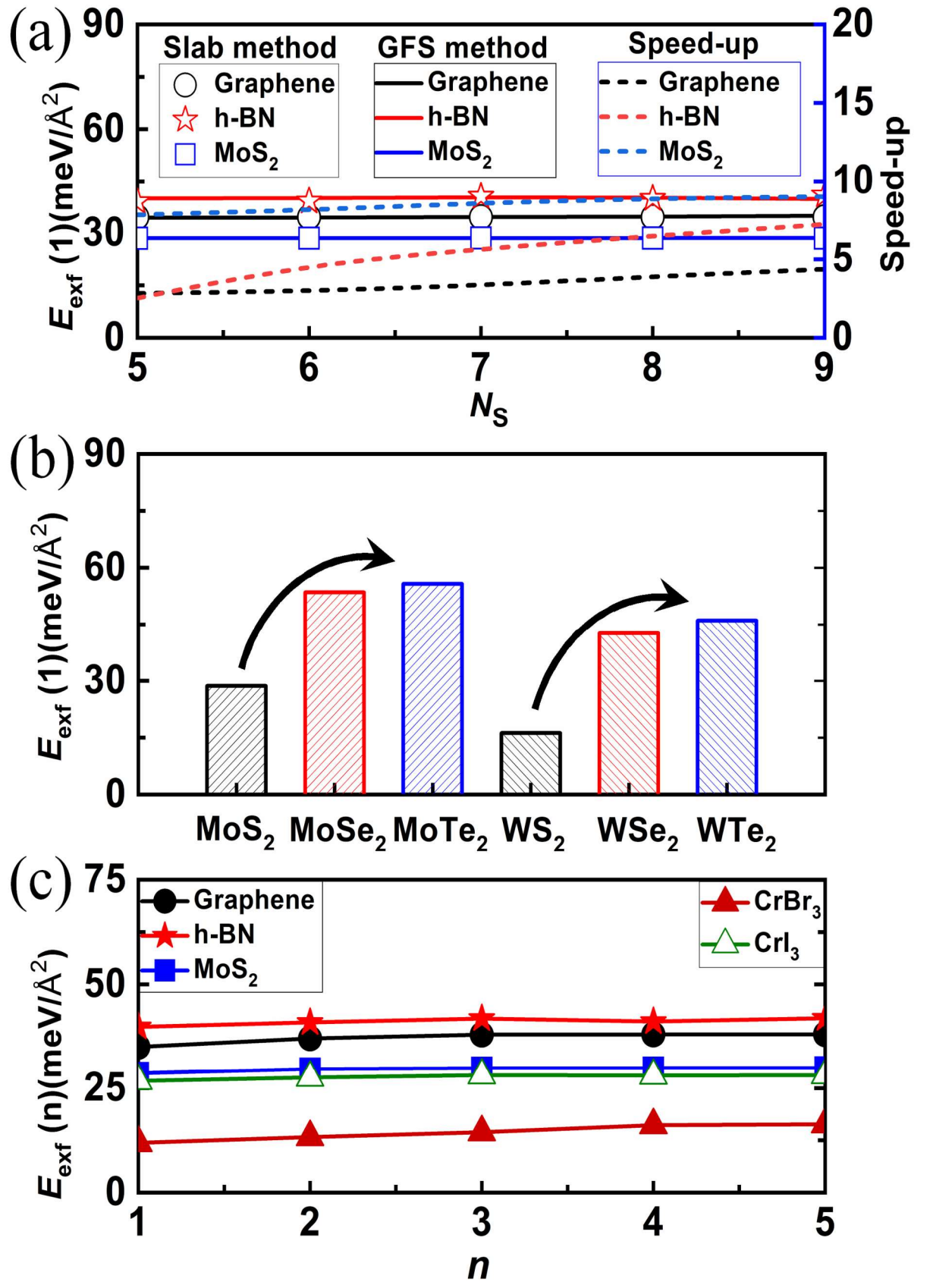

FIG. 2. (a) The monolayer exfoliation energies (per area) *versus* $N_{\mathrm{S}}$ of graphene, h-BN, and $MoS_2$ obtained by the traditional slab and Green's function surface methods, and the ratio of the required time for computing the monolayer exfoliation energy by the slab method to that by the GFS method. (b) The EEs of $MX_2$. (c) $n$-layer EEs of various 2D materials obtained by the GFS model.

The speed-up is more significant as the $N_{\mathrm{S}}$ increases. It is even as high as ten times for the $MoS_2$. Our calculated EEs of graphene, h-BN, and $MoS_2$ have the same trend to the plane-wave results [35], *i.e.*, EE(h-BN) $>$ EE(graphene) $>$ EE($MoS_2$). Moreover, it is found that the EEs of transition metal dichalcogenides $MX_2$ (M = Mo and W, X = S, Se, and Te) show the rules that EE($MoX_2$) $>$ EE($WX_2$) and EE($MS_2$) $<$ EE($MSe_2$) $<$ EE($MTe_2$), as shown in the Fig. 2(b).

We further evaluate the ($n > 1$) EEs of $n$-layer 2D materials by the GFS model. Taking graphene, h-BN, and $MoS_2$ as the samples, their $n$-layer EEs $E_{\mathrm{exf}}(\mathrm{n})$ can also be calculated by Eq. (2), while now $E_{\mathrm{II}}$ refers to the energy of exfoliated $n$-layer film. The $E_{\mathrm{exf}}(\mathrm{n})$(see Fig. 2(c)) are larger than their ML EEs $E_{\mathrm{exf}}(1)$ due to the stronger interaction with the bulk underneath, consistent with the recent report [35]. This indicates that it is easiest to exfoliate a ML from the bulk according to the exfoliation energy.

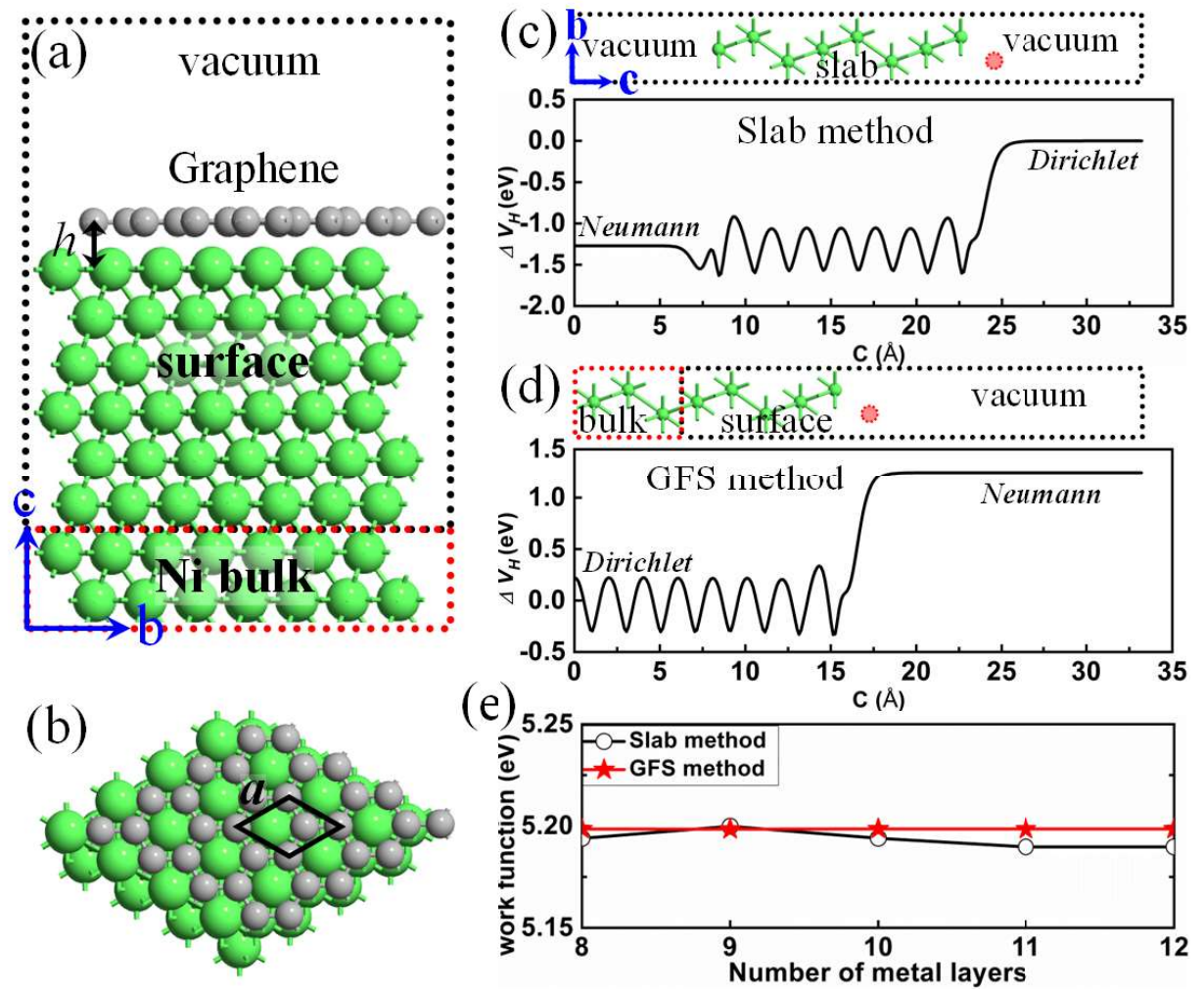

FIG. 3. (a) Side and (b) top views of graphene deposited on Ni(111) surface. Hartree difference potential obtained by (c) the slab and (d) Green's function surface methods. The red atoms of surface structures in (c) and (d) refer to the ghost atoms. (e) WFs *versus* the number of metal layers calculated with the slab and GFS methods.

Recently, the chromium trihalides $CrX_3$ (X = Cl, Br, and I) have become a hotspot due to their intrinsic magnetic semiconductor features [23–25], all of which exhibit a strong intralayer ferromagnetic coupling. While the interlayer coupling shows ferromagnetic for $CrBr_3$ and $CrI_3$, it is anti-ferromagnetic for $CrCl_3$[51]. Their MLs and few-layers can be prepared by means of mechanical exfoliation [23, 25]. In this work, we perform the spin-polarized calculations and obtain EEs of the interlayer ferromagnetic $CrBr_3$ and $CrI_3$ by the GFS model (see Fig. 2(c)).The EEs of $CrI_3$ are close to that of $MoS_2$, and twice as large as that of $CrBr_3$. This suggests that the $CrBr_3$ is more easily exfoliated than $CrI_3$ due to its weaker interlayer interaction.

Another approach widely adopted to prepare 2D materials is the CVD method. Taking graphene as an example, it is often grown on a Ni(111) surface because its lattice matches well with that of graphene [52, 53]. The C atoms are favorably adsorbed on the three-fold hollow sites of Ni(111) surface (see Figs. 3(a) and 3(b))[54]. Our optimized lattice parameter $a$ of graphene-Ni(111) system is 2.48 Å with a height ($h$) of 2.89 Å. In the following, we further investigate the exfoliation of monolayer and few-layer graphene from a semi-infinite Ni(111) surface.

Before calculating the EEs of graphene-Ni systems, we first demonstrate the effectiveness of the GFS approach for the determination of work function (WF) of Ni surface [55]. The WF is a fundamental feature of metal surfaces, and is defined as the difference between the invariable Hartree difference potential ($\Delta V_{\mathrm{H}}$) in the vacuum and the chemical potential $\mu$, *i.e.*, WF = $\Delta V_{\mathrm{H}} - \mu$. A ghost atom technique is adopted (see Figs. 3(c) and 3(d)) here and additional LCAO basis sets are positioned above the surface, to accurately describe the decay of surface charge densities into vacuum. One ghost atom is enough for the Ni (111) surface to obtain a convergent WF according to our test results (see Supplemental Material [34]). Note that the slab structure is a finite system and has a finite number of electrons. The total number of electrons change as transferring charge from a molecule to the surface (or the other way around). Hence, the chemical potential of the electrons in the slab also changes. Significantly, the GFS approach can completely alleviate this spurious effect through coupling the surface to an infinite electron reservoir (bulk region) at an invariable chemical potential.

Figures 3(c) and 3(d) show the basic difference between the GFS and slab models for WF calculations, as the GFS-calculated Ni(111) surface region is perfectly matched to its bulk region. The WF calculations are quite time-consuming for the slab model due to slow convergence *versus* the thick atomic layers. The Neumann and Dirichlet BCs are imposed on the left and right sides of the slab in the calculations, respectively. The $\Delta V_{\mathrm{H}}$ is zero in the vacuum on the right side of the slab (see Fig. 3(c)), and the slab-calculated WF is described as $\mathrm{WF}_{\mathrm{slab}} = -\mu_{\mathrm{slab}}$. For the GFS model, we adopt the Neumann and Dirichlet BCs in the vacuum and bulk-surface interface, respectively. On this occasion, the $\mu$ of the bulk region represents the entire surface system, thus the GFS-calculated WF is obtained by $\mathrm{WF}_{\mathrm{GFS}} = \Delta V_{\mathrm{H}} - \mu_{\mathrm{bulk}}$.

Figure 3(e) shows how the Ni(111) WF obtained by the slab (GFS) approach converges *versus* the atomic monolayer numbers in the slab (surface region). It demonstrates that rather thick slabs are required to converge the WF, but the GFS-calculated WF hardly depends on the thickness of surface region and converges fast to 5.20 eV, closer to the experimental data 5.35 eV. This is because the GFS-calculated surface states are well coupled to the bulk region, suggesting the bulk states are correctly taken into account in an accurate manner for the surface region with any thickness.

After obtaining a good description of the WF of Ni(111) surface using the GFS model, we next explore the exfoliation of ML and few-layer ($n$L) graphene from the same semi-infinite Ni substrate. These $n$L graphene plus Ni substrate systems are labeled as $n$L-Ni, as shown in Fig. 4. The results show a trend that the energy cost gradually decreases as the $n$-layer graphene is completely exfoliated from the same Ni substrate (see the green dashed line A). This is because more interlayer coupling of graphene weakens the graphene-Ni interface coupling, which is stronger than the interlayer coupling of graphene based on the higher 1L-Ni exfoliation energy. In addition, it is found that it is easiest to exfoliate a graphene ML (1L) as a tri-layer graphene (3L) is absorbed on the Ni substrate (see the red dotted line B). More importantly, we find a rule that it always has the lowest energy to exfoliate ($n$-1)-layer graphene from a $n$L-Ni system (see the blue circle). The main reason for this is that the graphene-Ni has a stronger interface interaction than that across graphene layers (see Fig. 2(a) and the 1L-Ni of Fig. 4). These findings suggest that one can obtain the free-standing graphene ML from Ni(111) surface through a two-step exfoliation approach. Namely, the first step is to exfoliate $n$-1 layers, and the second is to exfoliate the remaining monolayer.

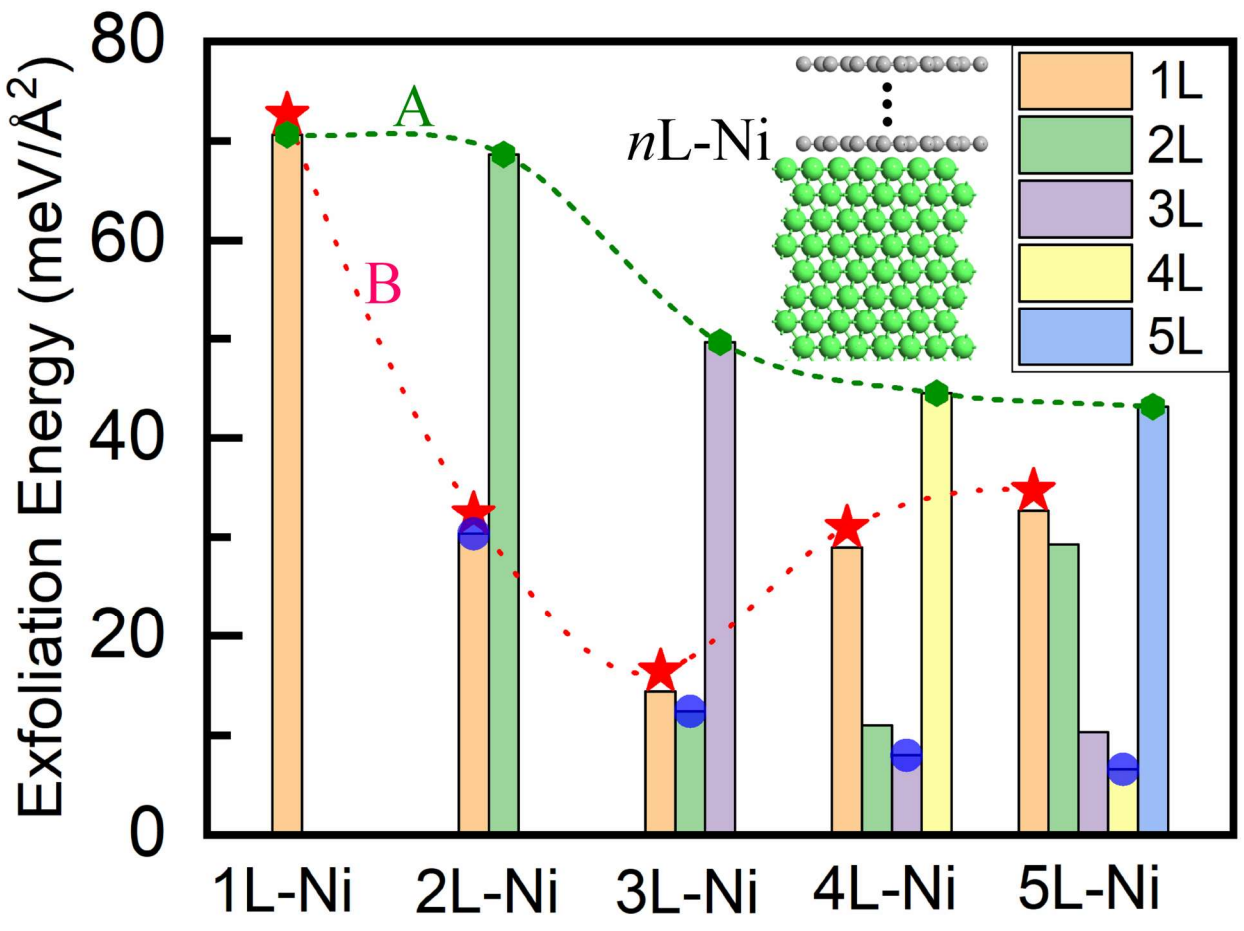

FIG. 4. The graphene-$n$L layer (from 1 to n) exfoliation energies (per area) of $n$L-Ni structures (see the inset).

Additionally, within the framework of GFS model, the angle-resolved photoemission spectroscopy can be well produced through calculating the spectral function of surface systems (see Supplemental Material [34]), which is one of the most important and direct methods of getting insights into the surface electronic structures of solids [56]. Some attempts have been made to simulate the ARPES, such as a one-step model based on the coherent potential approximation alloy theory [57, 58], and direct calculations from the energy-momentum dispersion and site-projected character [59]. Significantly, the DFT-based GFS approach can produce more reliable and clearer ARPES. In addition, the GFS model can deal with homogeneous crystals as well as complicated inhomogeneous structures, whereas the latter is difficult to obtain through the Wannier function approach [60–62].

The GFS-calculated ARPES of graphene-Ni systems are shown in Fig. 5. Our results of ARPES of the 1L-Ni (see Fig. 5(a)) show a good agreement with the experimental data [63]. Figures 5(b) and 5(c) show the ARPES of pristine Ni(111) and projected 1L-Ni in the graphene, respectively. The valence states of graphene are mostly submerged into the Ni substrate due to its stronger metallic nature. In addition, it is found that the conductive band moves towards the Fermi level($E_{\mathrm{F}}$) in the path of $\Gamma$-M($\mathrm{C}_{\Gamma\mathrm{M}}$) and has a pronounced bending along the M-K($\mathrm{C}_{\mathrm{MK}}$) route which causes a bandgap to appear for the graphene, according to its projected

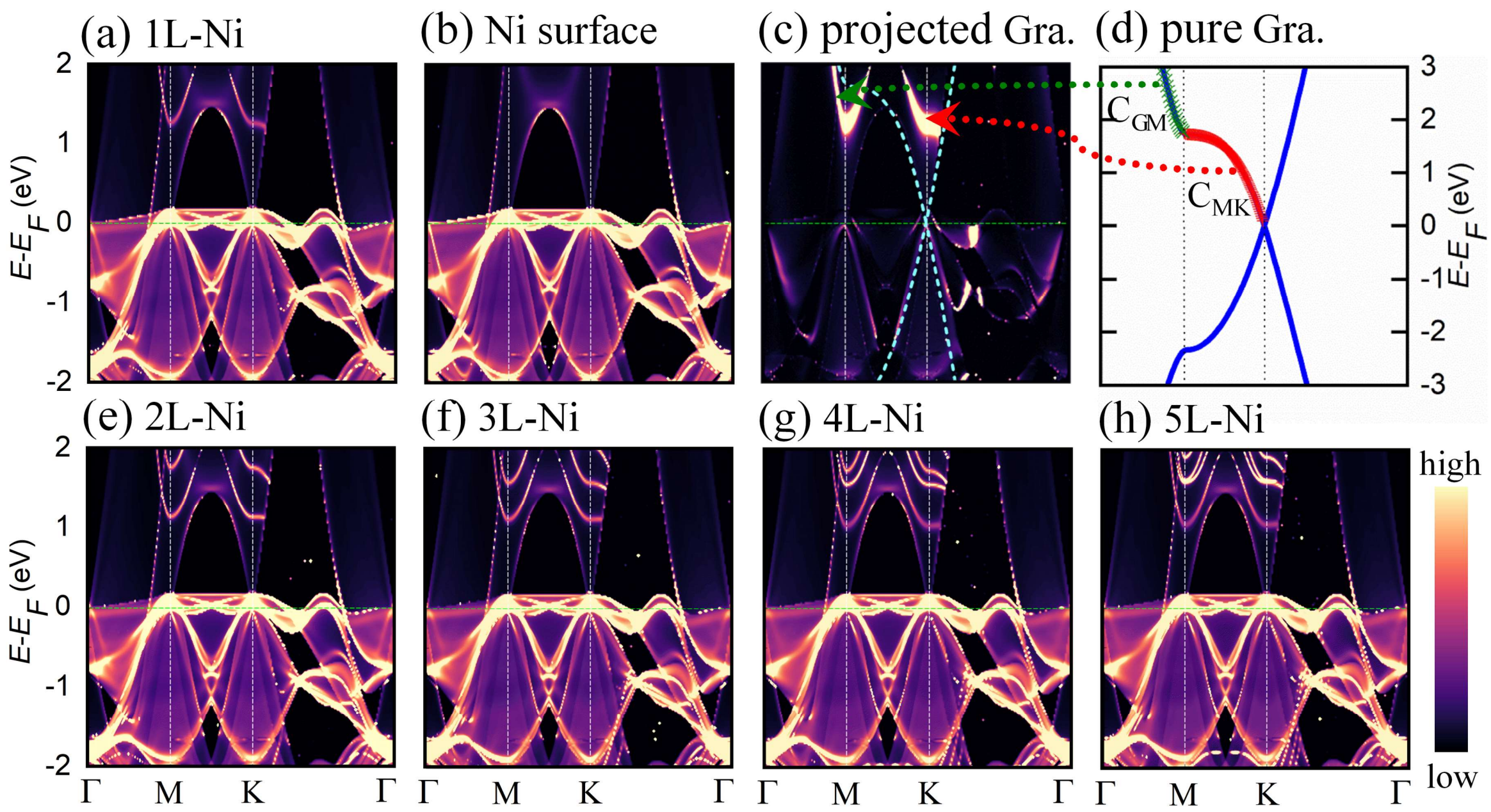

FIG. 5. ARPES of $n$L-Ni systems obtained by the Green's function surface approach. (a) 1L-Ni. (b) Pristine Ni(111) surface. (c) Projected ARPES of 1L-Ni in the graphene. (d) Pristine graphene band. (e)-(h) 2L-Ni to 5L-Ni. The $E_F$ is moved to zero.

ARPES (see Fig. 5(c)) and band structures (see Fig. 5(d)). This is because that the strong arched conduction states of Ni(111) surface (along M-K) cause the band bending of graphene. It demonstrates that the graphene and Ni(111) have a strong coupling, causing higher EE with respect to that from the graphene bulk. For the case of $n$L-Ni systems (see Figs. 5(e) to 5(h)), more conduction states of graphene can be observed, all of which move towards the $E_F$ in the path of Γ-M and bend along the M-K route due to the coupling with Ni substrate. These are consistent with their real-space local density of states projected on each surface structure (see Supplemental Material [34]), *i.e.*, a large energy gap appears above the $E_F$ for the projection of graphene due to the graphene-Ni interface coupling. Therefore, the GFS approach is one effective solution to produce the ARPES that can be compared directly to experimental data and unveil some important geometric and electronic structure features of certain 2D materials.

## IV. CONCLUSION

In summary, we propose a Green's function surface approach for studies of the exfoliation of ultrathin 2D layered materials either from their bulk crystals or from semi-infinite substrates. This approach has many significant advantages over previous methods. It includes, but is not limited to, the following aspects: 1) One can construct more realistic surface structures of exfoliating certain 2D material based on the GFS model, such as using a periodic unit-cell to describe its bulk crystal or thick sample. 2) It converges very fast with the number of atoms in model reduces computational cost. 3) More accurate work function of metal surface can be obtained by the GFS approach, and the exfoliation of 2D materials from semi-infinite substrates can be well described in a more realistic environment. Moreover, the GFS approach can be employed to produce the ARPES of surface systems, and is useful to examine the geometric and electronic structure features of 2D materials through comparison with experimental data. Our results demonstrate that the GFS model is powerful for studies of exfoliation issues and surface problems.

## V. ACKNOWLEDGEMENTS

The work at HNU was supported by the NSFC (Grants No. 11774079, No. 61774059, No. U1704136 and No. 61604061), the Scientific and Technological Innovation Program of Henan Province's Universities (Grant No. 20HASTIT026), the Young Backbone Teacher Program of Henan Province's Higher Education (Grant No. 2017GGJS043), the Henan Overseas Expertise Introduction Center for Discipline Innovation (Grant No. CXJD2019005), the Science Foundation for the Excellent Youth Scholars of HNU (Grant No. 2016YQ05), and the HPCC of HNU. The work at UCI was supported by the US DOE-BES (Grant No. DE-FG02-05ER46237). We thank W. Ju at HNUST for helpful discussion.

[1] K. S. Novoselov, A. K. Geim, S. V. Morozov, D. Jiang, Y. Zhang, S. V. Dubonos, I. V. Grigorieva, and A. A. Firsov, Science **306**, 666 (2004).

[2] L. Song, L. Ci, H. Lu, P. B. Sorokin, C. Jin, J. Ni, A. G. Kvashnin, D. G. Kvashnin, J. Lou, B. I. Yakobson *et al.*, Nano Lett. **10**, 3209 (2010).

[3] P. Vogt, P. De Padova, C. Quaresima, J. Avila, E. Frantzeskakis, M. C. Asensio, A. Resta, B. Ealet, and G. Le Lay, Phys. Rev. Lett. **108**, 155501 (2012).

[4] C.-H. Chen, W.-W. Li, Y.-M. Chang, C.-Y. Lin, S.-H. Yang, Y. Xu, and Y.-F. Lin, Phys. Rev. Appl. **10**, 044047 (2018).

[5] L. Li, Y. Yu, G. J. Ye, Q. Ge, X. Ou, H. Wu, D. Feng, X. H. Chen, and Y. Zhang, Nat. Nanotechnol. **9**, 372 (2014).

[6] R. Quhe, Q. Li, Q. Zhang, Y. Wang, H. Zhang, J. Li, X. Zhang, D. Chen, K. Liu, Y. Ye *et al.*, Phys. Rev. Appl. **10**, 024022 (2018).

[7] I. Pletikosić, F. von Rohr, P. Pervan, P. K. Das, I. Vobornik, R. J. Cava, and T. Valla, Phys. Rev. Lett. **120**, 156403 (2018).

[8] J. Zeng, P. Cui, and Z. Zhang, Phys. Rev. Lett. **118**, 046101 (2017).

[9] A. J. Mannix, X.-F. Zhou, B. Kiraly, J. D. Wood, D. Alducin, B. D. Myers, X. Liu, B. L. Fisher, U. Santiago, J. R. Guest *et al.*, Science **350**, 1513 (2015).

[10] B. Feng, J. Zhang, Q. Zhong, W. Li, S. Li, H. Li, P. Cheng, S. Meng, L. Chen, and K. Wu, Nat. Chem. **8**, 563 (2016).

[11] Y. An, Y. Hou, H. Wang, J. Li, R. Wu, T. Wang, H. Da, and J. Jiao, Phys. Rev. Appl. **11**, 064031 (2019).

[12] Y. An, J. Jiao, Y. Hou, H. Wang, R. Wu, C. Liu, X. Chen, T. Wang, and K. Wang, J. Phys.: Condens. Matter **31**, 065301 (2019).

[13] Q. H. Wang, K. Kalantar-Zadeh, A. Kis, J. N. Coleman, and M. S. Strano, Nat. Nanotech. **7**, 699 (2012).

[14] M. Choi, Phys. Rev. Appl. **9**, 024009 (2018).

[15] D. Lagarde, L. Bouet, X. Marie, C. R. Zhu, B. L. Liu, T. Amand, P. H. Tan, and B. Urbaszek, Phys. Rev. Lett. **112**, 047401 (2014).

[16] C. Wang, B. Lian, X. Guo, J. Mao, Z. Zhang, D. Zhang, B.-L. Gu, Y. Xu, and W. Duan, Phy. Rev. Lett. **123**, 126402 (2019).

[17] B. Anasori, M. R. Lukatskaya, and Y. Gogotsi, Nat. Rev. Mater. **2**, 16098 (2017).

[18] X.-H. Zha, J. Zhou, K. Luo, J. Lang, Q. Huang, X. Zhou, J. S. Francisco, J. He, and S. Du, J. Phys.: Condens. Matter **29**, 165701 (2017).

[19] F.-f. Zhu, W.-j. Chen, Y. Xu, C.-l. Gao, D.-d. Guan, C.-h. Liu, D. Qian, S.-C. Zhang, and J.-f. Jia, Nat. Mater. **14**, 1020 (2015).

[20] H. Zhang, J. Xiong, M. Ye, J. Li, X. Zhang, R. Quhe, Z. Song, J. Yang, Q. Zhang, B. Shi *et al.*, Phys. Rev. Appl. **11**, 064001 (2019).

[21] J. Zhang, Y. Chen, and X. Wang, Energ. Environ. Sci. **8**, 3092 (2015).

[22] J. Zhang, M. Deng, Y. Yan, T. Xiao, W. Ren, and P. Zhang, Phys. Rev. Appl. **11**, 044052 (2019).

[23] B. Huang, G. Clark, E. Navarro-Moratalla, D. R. Klein, R. Cheng, K. L. Seyler, D. Zhong, E. Schmidgall, M. A. McGuire, D. H. Cobden *et al.*, Nature **546**, 270 (2017).

[24] D. R. Klein, D. MacNeill, J. L. Lado, D. Soriano, E. Navarro-Moratalla, K. Watanabe, T. Taniguchi, S. Manni, P. Canfield, J. Fernández-Rossier *et al.*, Science **360**, 1218 (2018).

[25] B. Huang, G. Clark, D. R. Klein, D. MacNeill, E. Navarro-Moratalla, K. L. Seyler, N. Wilson, M. A. McGuire, D. H. Cobden, D. Xiao *et al.*, Nat. Nanotech. **13**, 544 (2018).

[26] D.-H. Kim, K. Kim, K. -T. Ko, J. Seo, J. S. Kim, T.-H. Jang, Y. Kim, J.-Y. Kim, S.-W. Cheong, and J.-H. Park, Phys. Rev. Lett. **122**, 207201 (2019).

[27] Y. Deng, Y. Yu, Y. Song, J. Zhang, N. Z. Wang, Z. Sun, Y. Yi, Y. Z. Wu, S. Wu, J. Zhu *et al.*, Nature **563**, 94 (2018).

[28] B. Feng, R.-W. Zhang, Y. Feng, B. Fu, S. Wu, K. Miyamoto, S. He, L. Chen, K. Wu, K. Shimada *et al.*, Phys. Rev. Lett. **123**, 116401 (2019).

[29] V. Nicolosi, M. Chhowalla, M. G. Kanatzidis, M. S. Strano, and J. N. Coleman, Science **340**, 1226419 (2013).

[30] Y. Zhu, S. Murali, W. Cai, X. Li, J. W. Suk, J. R. Potts, and R. S. Ruoff, Adv. Mater. **22**, 3906 (2010).

[31] M. Naguib, M. Kurtoglu, V. Presser, J. Lu, J. Niu, M. Heon, L. Hultman, Y. Gogotsi, and M. W. Barsoum, Adv. Mater. **23**, 4248 (2011).

[32] A. Reina, X. Jia, J. Ho, D. Nezich, H. Son, V. Bulovic, M. S. Dresselhaus, and J. Kong, Nano Lett. **9**, 30 (2009).

[33] C. Tan and H. Zhang, Nat. Commun. **6**, 7873 (2015).

[34] See Supplemental Material at http://link.aps.org/supplemental/XXX for (A) the traditional slab model and (B) GFS model for calculating the exfoliation energies, (C)test of the ghost atom, (D) the GFS-calculated ARPES, and (E) projected local density of states of nL-Ni surface structures.

[35] J. H. Jung, C.-H. Park, and J. Ihm, Nano Lett. **18**, 2759 (2018).

[36] N. Papior, N. Lorente, T. Frederiksen, A. García, and M. Brandbyge, Comput. Phys. Commun. **212**, 8 (2017).

[37] S. Smidstrup, D. Stradi, J. Wellendorff, P. A. Khomyakov, U. G. Vej-Hansen, M.-E. Lee, T. Ghosh, E. Jónsson, H. Jónsson, and K. Stokbro, Phys. Rev. B **96**, 195309 (2017).

[38] J. Taylor, H. Guo, and J. Wang, Phys. Rev. B **63**, 121104(R) (2001).

[39] M. Brandbyge, J.-L. Mozos, P. Ordejón, J. Taylor, and K. Stokbro, Phys. Rev. B **65**, 165401 (2002).

[40] J. M. Soler, E. Artacho, J. D. Gale, A. García, J. Junquera, P. Ordejón, and D. Sánchez-Portal, J. Phys.: Condens. Matter **14**, 2745 (2002).

[41] S. Smidstrup, T. Markussen, P. Vancraeyveld, J. Wellendorff, J. Schneider, T. Gunst, B. Verstichel, D. Stradi, P. A. Khomyakov, U.G. Vej-Hansen *et al.*, J. Phys.: Condens. Matter **32**, 015901 (2020).

[42] J. P. Perdew, J. A. Chevary, S. H. Vosko, K. A. Jackson, M. R. Pederson, D. J. Singh, and C. Fiolhais, Phys. Rev. B **46**, 6671 (1992).

[43] J. P. Perdew, K. Burke, and M. Ernzerhof, Phys. Rev. Lett. **77**, 3865 (1996).

[44] D. R. Hamann, Phys. Rev. B **88**, 085117 (2013).

[45] S. Grimme, J. Comput. Chem. **27**, 1787 (2006).

[46] E. Ziambaras, J. Kleis, E. Schröder, and P. Hyldgaard, Phys. Rev. B **76**, 155425 (2007).

[47] Y. Jing, Y. Ma, Y. Li, and T. Heine, Nano Lett. **17**, 1833 (2017).
[48] F. Li, X. Liu, Y. Wang, and Y. Li, J. Mater. Chem. C **4**, 2155 (2016).
[49] D. Stradi, U. Martinez, A. Blom, M. Brandbyge, and K. Stokbro, Phys. Rev. B **93**, 155302 (2016).
[50] W. Wang, S. Dai, X. Li, J. Yang, D. J. Srolovitz, and Q. Zheng, Nat. Commun. **6**, 7853 (2015).
[51] R. W. Bené, Phys. Rev. **178**, 497 (1969).
[52] S. Helveg, C. López-Cartes, J. Sehested, P. L. Hansen, B. S. Clausen, J. R. Rostrup-Nielsen, F. Abild-Pedersen, and J. K. Nørskov, Nature **427**, 426 (2004).
[53] A. Dahal and M. Batzill, Nanoscale **6**, 2548 (2014).
[54] Z. Fu and Y. An, RSC Adv. **6**, 91157 (2016).
[55] G. Giovannetti, P. A. Khomyakov, G. Brocks, V. M. Karpan, J. van den Brink, and P. J. Kelly, Phys. Rev. Lett. **101**, 026803 (2008).
[56] T. Shoman, A. Takayama, T. Sato, S. Souma, T. Takahashi, T. Oguchi, K. Segawa, and Y. Ando, Nat. Commun. **6**, 6547 (2015).
[57] J. Minár, J. Braun, S. Mankovsky, and H. Ebert, J. Electron Spectrosc. **184**, 91 (2011).
[58] G. Derondeau, J. Braun, H. Ebert, and J. Minár, Phys. Rev. B **93**, 144513 (2016).
[59] C. D. Spataru and F. Léonard, Phys. Rev. B **90**, 085115 (2014).
[60] N. Marzari, A. A. Mostofi, J. R. Yates, I. Souza, and D. Vanderbilt, Rev. Mod. Phys. **84**, 1419 (2012).
[61] H. Zhang, C.-X. Liu, X.-L. Qi, X. Dai, Z. Fang, and S.-C. Zhang, Nat. Phys. **5**, 438 (2009).
[62] X.-L. Sheng and B. K. Nikolić, Phys. Rev. B **95**, 201402(R) (2017).
[63] Y. S. Park, J. H. Park, H. N. Hwang, T. S. Laishram, K. S. Kim, M. H. Kang, and C. C. Hwang, Phys. Rev. X **4**, 031016 (2014).